\documentclass[aps,pra,letterpaper,twocolumn,floatfix]{revtex4-2}
\usepackage{amsmath,amssymb} 
\usepackage[T1]{fontenc}
\usepackage[utf8]{inputenc} 

\usepackage{graphicx} 

\usepackage{hyperref}
\hypersetup{colorlinks=true,citecolor={black},linkcolor={blue},urlcolor={blue}} 
\usepackage{xcolor}
\usepackage{bm}

\DeclareMathOperator{\sn}{sn}
\DeclareMathOperator{\cn}{cn}
\DeclareMathOperator{\dn}{dn}

\newcommand{\cPT}{\ensuremath{\mathcal{PT}}}
\newcommand{\hs}[1]{\hat{s}_{#1}}
\newcommand{\mi}{\mathrm{i}}
\newcommand{\ph}[1]{\hat{\psi}^{\vphantom{\dag}}_{#1}}
\newcommand{\pd}[1]{\hat{\psi}^\dag_{#1}}
\newcommand{\phh}[1]{\hat{\psi}^{2\vphantom{\dag}}_{#1}}
\newcommand{\pdd}[1]{\hat{\psi}^{\dag2}_{#1}}
\newcommand{\ve}{\varepsilon}

\newcommand{\snt}{\sn({\omega}t,m)}
\newcommand{\cnt}{\cn({\omega}t,m)}
\newcommand{\dnt}{\dn({\omega}t,m)}

\graphicspath{{.}{./figs/}}

\begin{document}

\title{Polarization dynamics of trapped polariton condensates with \cPT-symmetry}

\author{I. Jesán Velázquez-Reséndiz} 
\affiliation{Instituto de Energías Renovables, Universidad Nacional Autónoma de México, Temixco, Morelos, 62580, Mexico}

\author{Yuri G. Rubo} 
\email{ygr@ier.unam.mx}
\affiliation{Instituto de Energías Renovables, Universidad Nacional Autónoma de México, Temixco, Morelos, 62580, Mexico} 

\begin{abstract}
We propose a grated microcavity setup to form trapped polariton condensates with parity-time (\cPT) symmetry and study their polarization dynamics. The pseudo-conservative dynamics of the Stokes vector in proposed configuration is preserved in the presence of polariton-polariton interaction.
In the case of weak gain-dissipation inbalance, as compared to the linear polarization splitting, the polarization Stokes spheres are deformed into ellipsoids. 
When linear polarization splitting becomes weak, i.e., when \cPT-symmetry is broken for noninteracting system, the Stokes spheres are transformed into hyperboloids, but the dynamics is still described by closed trajectories, allowing manipulation of the polarization of polariton condensate by changing polarization splitting without losing its coherency.
\end{abstract}

\date{\today}

\maketitle

\emph{Introduction.} 
Discovery of exciton-polariton condensation and lasing \cite{kasprzak06,balili07} marked an important milestone in modern nonlinear optics. Due to the excitonic component of polariton quasiparticle, the space configuration of the condensates can be managed by pump and external fields, while the polarization of the condensate provides evidences of spontaneous symmetry breaking and monitors the dynamics of the order parameter \cite{baumberg08,levrat10,colas15}. 
Even in the simplest case of single trapped polariton condensate, which is spatially disconnected from the reservoirs of incoherent exciton-polariton created by pumping \cite{askitopoulos13,cristofolini13}, the polarization state can be rather nontrivial. It can exhibit formation of circular polarization degree with random handedness, manifesting spontaneous breaking of parity \cite{ohadi15}. Furthermore, the handedness can be manipulated by applied electric field, which controls the splitting $\ve$ between X and Y linearly polarized states, and this can be useful for computation with low energy consumption \cite{dreismann16}. 

While the polariton condensate is driven-dissipative system, the symmetry breaking resulting in
spontaneous formation of circular polarization could be understood by considering the condensate as a conservative Hamiltonian system. In this approximation, the dynamics is mapped to that of the Bose-Hubbard dimer \cite{smerzi97,milburn97,franzosi01,albiez05,zibold10}, or, if one uses the Schwinger realization of the angular momentum, to the dynamics of Lipkin-Meshkov-Glick model \cite{lipkin65,vardi01}, where the parity symmetry breaking is known as formation of self-trapped many-body states \cite{holthaus01,pudlik14}. 
The behavior of these systems, especially close to the long-period classical trajectories, has attracted much attention recently to study scrambling \cite{xu20,pilatowskycameo20}, quantum phase transitions \cite{wang19}, and possible applications for classical and quantum informatics. 

Based on these features of polariton dimers, there is growing interest recently in use of polariton condensates as possible elements for quantum computing, similarly to how it is done with the superconducting qubits \cite{arute19}.
The main obstacle comes from the fact the polariton condensates are dissipative systems, and to produce and maintain them it is necessary to apply external pumping, which easily destroys long-time coherence and related quantum effects like entanglement between the condensates in a network. 
To apply polariton condensates for information and computation goals it is desirable to force the condensate to possess conservative or pseudo-conservative dynamics, which is characterized by closed orbits in classical phase space, or by the real energies in quantum limit. A promising method to achieve this is to form parity-time (\cPT) symmetric polariton condensates \cite{chestnov21,kalozoumis21}. 

Since the pioneering work by Bender and Boettcher \cite{bender98}, the non-Hermitian \cPT-symmetric systems have attracted much attention both theoretically and experimentally, see \cite{bender19book} for a review. These systems can possess real-valued energy spectrum if the \cPT-symmetry is unbroken, or pairs of complex-conjugate energy eigenvalues when the symmetry is spontaneously broken. The latter case is usually achieved for big enough non-Hermitian part of the Hamiltonian. In the typical example of a two-state quantum system, characterized by the coherent coupling $\ve$ and dissipative coupling $\gamma$ between the states, the \cPT-symmetry becomes broken for $|\gamma|>|\ve|$, which is unwanted if one aims to manipulate the system state by adiabatic change of $\ve$. The account for polariton-polariton interaction leads to non-linear equations with even more complex behavior \cite{konotop16}. In particular, the spontaneous parity breaking in the case of Bose-Hubbard dimer leads also to suppression of time reversal \cite{sukhorukov10}. This results in broken \cPT-symmetry even for small dissipative coupling parameter, that is  manifested by the blow-up dynamics \cite{barashenkov13} with disclosed classical trajectories. 

To overcome these difficulties, in this paper we propose and analyze a special grated microcavity setup, that introduces dissipation inbalance and polarization splitting between different linearly polarized states. This allows building the \cPT-symmetric polariton condensates, and we demonstrate the persistence of pseudo-conservative dynamics even at high occupation levels, where polariton-polariton interaction and related nonlinearities are important.

\begin{figure}[t]
	\includegraphics[width=0.46
    \textwidth]{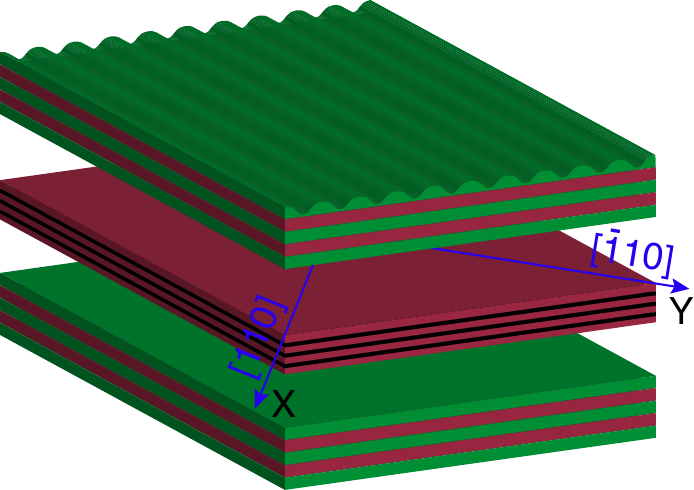} 
	\caption{
	Schematic view of the microcavity with sub-wavelength grating of one distributed Bragg mirror. 
	The grating induces different dissipation rates for polaritons with diagonal [100] and antidiagonal [010] linear polarizations,
	while there is energy splitting between [110] (X direction) and [$\bar{1}$10] (Y direction) linearly polarized polaritons. 
	}
    \label{fig:1}
\end{figure}

\emph{Microcavity structure and the model.} 
The polaritons in semiconductor microcavities present linear polarization dichroism, characterized by the splitting $\ve$ between X (horizontal) and Y (vertical) linear polarizations, that are defined by the crystallographic [110] and [$\bar{1}$10] axes, as shown in Fig.\ \ref{fig:1}.
This splitting appears mainly due to the mixture of the light- and the heavy-hole exciton components of polariton wave function on the low symmetry interfaces of the quantum wells \cite{aleiner92,ivchenko96,malpuech06}. The splitting is typically about tens of {\textmu}eV. 

Apart from the Hermitian splitting between the linear polarization components, we propose to introduce polarization-dependent dissipation. This can be achieved by weak sub-wavelength grating (SWG) of the microcavity surface, which makes reflectivity of the distributed Bragg mirror dependent on the polariton polarization \cite{huang07,kim16}.     
To obtain suitable dissipation inbalance one can use microcavity grated along diagonal or antidiagonal direction, i.e., along either [100] or [010] direction. This introduces a difference in the lifetimes of polaritons with diagonal and antidiagonal polarizations. In this configuration, the single-polariton Hamiltonian describing polarization state of a trapped polariton can be written in circular polarization basis as
\begin{equation}\label{Hams}
	H_s=-\frac{1}{2}\begin{pmatrix}{\mi}g & \ve-\mi\ve^\prime+\gamma \\ \ve+\mi\ve^\prime-\gamma & {\mi}g \end{pmatrix}.
\end{equation}
Here $g=\Gamma-W$, where $W$ is the external non-resonant pumping rate, $\Gamma$ is the average dissipation rate, and $\gamma$ defines the dissipation inbalance (we set $\hbar=1$). In general, weak SWG induces an addition splitting, which we encoded in Eq.\ \eqref{Hams} by the parameter $\ve^\prime$. It is important to note that the Hermitian parts of the single-polariton Hamiltonian \eqref{Hams} can be tuned by applied electric field \cite{dreismann16} and by the strain \cite{klopotowski06,balili10}, and in what follows, we assume the additional splitting is removed, $\ve^\prime=0$. 
As a result, when external pump of the condensate matches the average dissipation from the microcavity, i.e., $g=0$, the condensate is described by the \cPT-symmetric Hamiltonian $H_{s0}=-(\ve\sigma_x+\mi\gamma\sigma_y)/2$ with the energies $\pm\sqrt{\ve^2-\gamma^2}/2$. The general description of the $2\times2$ \cPT-symmetric matrices can be found in Ref.\ \cite{wang13}. In our case, the parity operation is defined by the Pauli $x$-matrix, $\mathcal{P}=\sigma_x$, while the time-reversal operation is $\mathcal{T}=K\sigma_x$, where $K$ is the complex-conjugation, so that any real matrix is in fact \cPT-symmetric. It is worth noting that the Hamiltonian \eqref{Hams} describes as well asymmetric hoping between two localized states, which for the case of a chain of such states is known as the Hatano-Nelson model \cite{hatano96}. 

The many-body extension of our model accounts for interaction of polaritons with the same circular polarization, which corresponds to the following equations
\begin{equation}\label{HeisEqs}
	\mi\frac{d\ph{\pm1}}{dt}=-\frac{1}{2}(\ve\pm\gamma)\ph{\mp1}+\frac{\alpha}{2}\pd{\pm1}\hat{\psi}_{\pm1}^2,
\end{equation}
for the dynamics of annihilations operators of polaritons with the right ($+1$) and the left ($-1$) circular polarizations, where $\alpha$ is the interaction constant. 
It is convenient to introduce the spin operators as $\hs{k}=\frac{1}{2}\left(\Psi^\dag\cdot\sigma_k\cdot\Psi\right)$ for $k=0,x,y,z$, where the column vector $\Psi=(\ph{+1},\ph{-1})^\mathsf{T}$, $\sigma_{x,y,z}$ are the Pauli matrices, and $\sigma_0$ is the $2\times2$ identity matrix. The spin dynamics is then governed by the equations
\begin{subequations}\label{SOperEqs}
\begin{align}
	& \frac{d\hs{0}}{dt}=-{\gamma}\hs{y}, \quad \frac{d\hs{x}}{dt}=-\frac{\alpha}{2}\{\hs{z},\hs{y}\}, \\
	& \frac{d\hs{y}}{dt}=-{\gamma}\hs{0}+{\ve}\hs{z}+\frac{\alpha}{2}\{\hs{z},\hs{x}\}, \quad 
	\frac{d\hs{z}}{dt}=-{\ve}\hs{y},
\end{align}
\end{subequations}
where $\{\hs{k},\hs{l}\}=\hs{k}\hs{l}+\hs{l}\hs{k}$.

It is easy to see that the single-polariton energies are real and the \cPT-symmetry is unbroken for $|\ve|>|\gamma|$, and it is important to note that the presence of polariton-polariton interaction does not change this fact. The non-Hermitian Bose-Hubbard Hamiltonian that corresponds to Eqs.\ \eqref{HeisEqs} is 
\begin{multline}\label{HamOp}
	\hat{\mathcal{H}}=-\frac{(\ve+\gamma)}{2}\pd{+1}\ph{-1}-\frac{(\ve-\gamma)}{2}\pd{-1}\ph{+1} \\
	+\frac{\alpha}{4}[\pdd{+1}\phh{+1}+\pdd{-1}\phh{-1}],
\end{multline}
or, using the spin operators,
\begin{subequations}\label{HamS}
\begin{align}
	& \mathcal{H}=H_0(\hat{\mathbf{s}})+{\mi}H_1(\hat{\mathbf{s}}), \\
	& H_0(\hat{\mathbf{s}})=-\ve\hs{x}+\frac{\alpha}{2}[\hat{s}_z^2+\hat{s}_0^2-\hat{s}_0], \\ 
	& H_1(\hat{\mathbf{s}})=-\gamma\hs{y}.
\end{align}
\end{subequations}
This Hamiltonian can be transformed into an Hermitian operator by the Dyson transformation $e^{-t\hat{s}_z}\mathcal{H}e^{t\hat{s}_z}$ with $\tanh(t)=\gamma/\ve$, which eliminates the $H_1$ part and renormalizes the Josephson coupling parameter to $\tilde{\ve}=\sqrt{\ve^2-\gamma^2}$. 

In the opposite case, $|\ve|<|\gamma|$, the \cPT-symmetry is broken and the single-polariton energies becomes complex. Including the polariton-polariton interaction actually improves the situation. As we show below in the mean-field approximation valid for large occupation numbers, this model preserves pseudo-conservative dynamics even for small splittings $\ve$.

\emph{Semiclassical dynamics.}
In the mean-field approximation, the spin operator $\hat{\mathbf{s}}=\{\hs{x},\hs{y},\hs{z}\}$ is replaced by a 3D vector $\mathbf{S}$ with the length $S{\equiv}S_0$. The length of the spin is not conserved in our case and it satisfies the dynamical equation 
\begin{equation}\label{VecSeq}
	\frac{d\mathbf{S}}{dt}=\left[\frac{dH_0}{d\mathbf{S}}\times\mathbf{S}\right]+S\frac{dH_1}{d\mathbf{S}},
\end{equation}	
or in the components
\begin{subequations}\label{CompSeqs}
\begin{alignat}{2}
	& \dot{S}_x=-{\alpha}S_zS_y, \quad && \dot{S}_y=-{\gamma}S+{\ve}S_z+{\alpha}S_zS_x, \\
	& \dot{S}_z=-{\ve}S_y, \quad && \dot{S}=-{\gamma}S_y.
\end{alignat}
\end{subequations}

In what follows, we will assume that the parameters $\alpha$, $\gamma$, and $\ve$ have positive values. In the case of negative values, one can reestablish the equations (\ref{CompSeqs}a,b) with positive parameters by using appropriate transformations of spin components. E.g., in the case of attractive interaction, $\alpha<0$, we apply $S_x\,\rightarrow\,-S_x$. In the case of $\ve<0$ we apply $S_z\,\rightarrow\,-S_z$ together with $S_x\,\rightarrow\,-S_x$. Finally, in the case of $\gamma<0$ we invert $S_z\,\rightarrow\,-S_z$ and $S_y\,\rightarrow\,-S_y$. By choosing $\ve^{-1}$ as the unit of time we can always set $\ve=1$, and we also can rescale the spin $\alpha\mathbf{S}=\mathbf{s}=\{x,y,z\}$ to obtain 
\begin{subequations}\label{XYZset}
\begin{alignat}{2}
	& \dot{x}=-zy, \quad && \dot{y}=-{\gamma}s+z+zx, \\
	& \dot{z}=-y, \quad && \dot{s}=-{\gamma}y.
\end{alignat}
\end{subequations}

The system of equations \eqref{XYZset} has two integrals of motion, the ``energy'' $E$ and the parameter $\rho$, that defines the spin size
\begin{equation}\label{E&rho}
    E=-x+\frac{1}{2}z^{2}, \qquad \rho=s-{\gamma}z.
\end{equation}
Applying these invariants we obtain the following equation for the $z$-component
\begin{align}\label{DifEq4z}
    \left(\frac{dz}{dt}\right)^2&=\left(\rho+{\gamma}z\right)^{2}-\left(\frac{z^{2}}{2}-E\right)^2-z^{2} \nonumber\\
    &=-\frac{1}{4}(z-z_{1})(z-z_{2})(z-z_{3})(z-z_{4}),
\end{align}
where in stead of the parameters $E$, $\rho$, $\gamma$ we also defined the four roots $z_{1,2,3,4}$ of the quartic polynomial in the right-hand side. They are subject to $z_1+z_2+z_3+z_4=0$.  

Eq.\ \eqref{DifEq4z} can be solved analytically using a Möbius (homographic) transformation 
\begin{equation}\label{Mobius}
	w(t)=\frac{az(t)+b}{cz(t)+d}
\end{equation} 
and adjusting the constants $a$, $b$, $c$, $d$ to obtain the equation for the Jacobi elliptic cosine function for $w(t)$, similar to discussion in Ref.\ \cite{schwalm15}. In our case, special care should be taken concerning the number of real roots, since one can have either all four real roots $z_{1,2,3,4}$ or only two real roots and a pair of complex ones. 
As a result, the solution of \eqref{DifEq4z} can be written as 
\begin{equation}\label{Zsol}
	z(t)=\frac{\omega_1z_4[1\pm\cnt]+\omega_4z_1[1\mp\cnt]}{\omega_1[1\pm\cnt]+\omega_4[1\mp\cnt]},
\end{equation}
where 
\begin{subequations}\label{NUsMOmega}
\begin{align}
    & \omega_{1}=\sqrt{(z_{1}-z_{2})(z_{1}-z_{3})}, \quad \omega_{4}=\sqrt{(z_{3}-z_{4})(z_{2}-z_{4})}, \\
	& \omega=\frac{1}{2}\sqrt{\omega_1\omega_4}, \qquad          		m=\frac{[\omega_1\omega_4+(z_1-z_2)(z_3-z_4)]^2}{4\omega_1\omega_4(z_1-z_2)(z_3-z_4)}.
\end{align}
\end{subequations}
When there are only two real roots, they are chosen as $z_1{\le}z_4$, while $z_{2}=z_{3}^{*}$. The modulus of elliptic cosine function $m<1$ in this case, $\cnt$ oscillates between $\pm1$, and the upper and lower signs in \eqref{Zsol} describe the same solution, which oscillates between $z_1$ and $z_4$. On the other hand, when there are four real roots, they are assumed to be ordered as $z_{1}{\leq}z_{2}{\leq}z_{3}{\leq}z_{4}$. In this case, the modulus $m>1$ and $\cnt=\dn(\sqrt{m}\,{\omega}t,1/m)$ oscillates between 1 and $\sqrt{m-1/m}$, so that the upper sign in \eqref{Zsol} describes oscillations between $z_3$ and $z_4$, while the lower sign corresponds to oscillations between $z_1$ and $z_2$. Finally, knowing $z(t)$ one can calculate $x(t)=(1/2)z(t)^2-E$, and using $y=-\dot{z}$ obtain
\begin{equation}\label{Ysol}
	y(t)=\pm\frac{8\omega^{3}(z_{4}-z_{1})\dnt\snt}{\left(\omega_1[1\pm\cnt]+\omega_4[1\mp\cnt]\right)^{2}}.
\end{equation}

\begin{figure}[t]
	\includegraphics[width=0.48
    \textwidth]{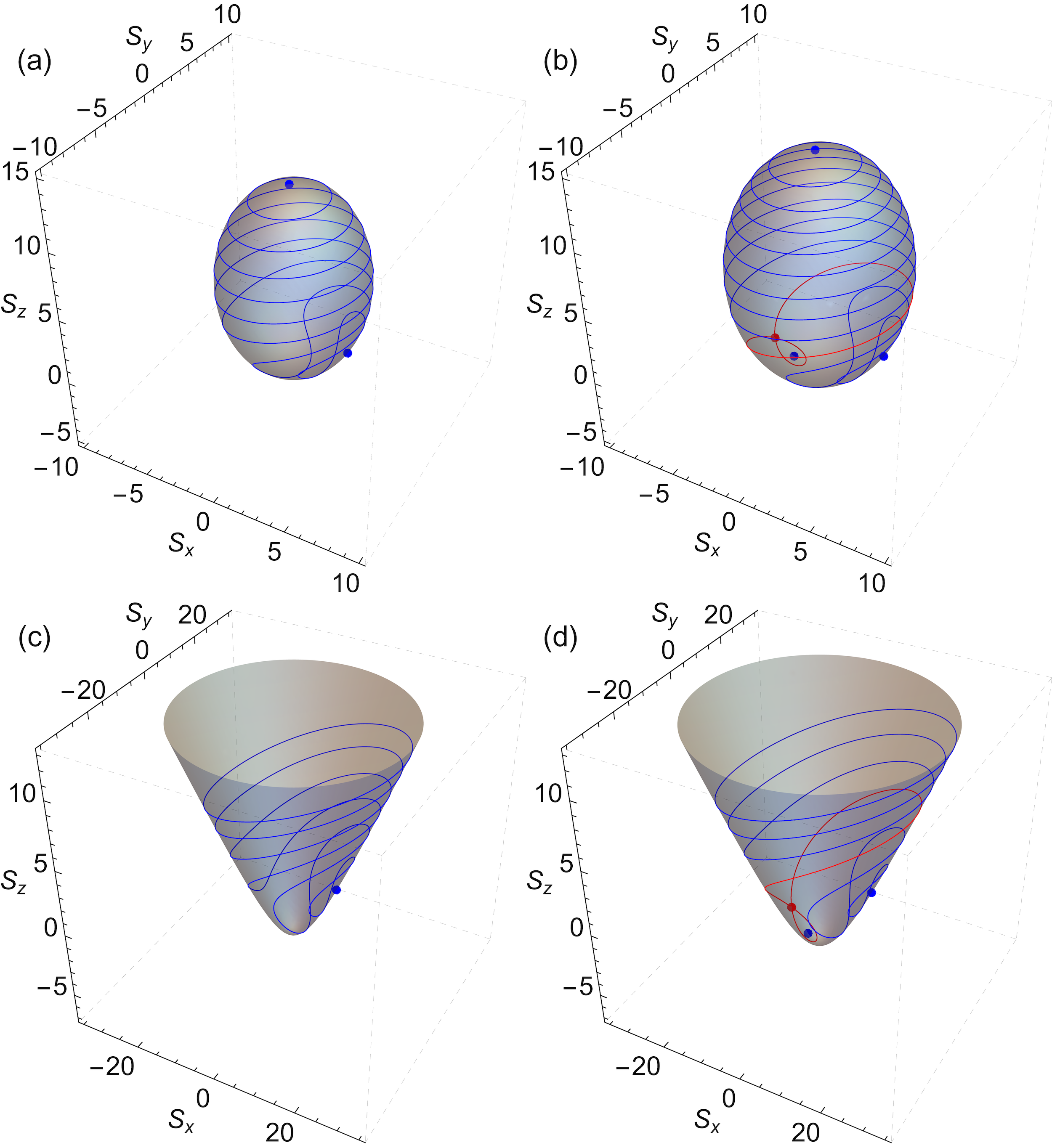} 
	\caption{
	Different topologies of the manifolds of spin trajectories. 
	Showing weak dissipative coupling cases with $\gamma/\ve=0.7$ that leads to $\rho_{c}=3.741$, 
	with $\rho=3.6$ (a) and $\rho=4.3$ (b), corresponding to the spin moving on ellipsoids; and 
	strong dissipative coupling cases with $\gamma=1.6$ that gives $\rho_{c}=8.729$, 
	with $\rho=8.5$ (c) and $\rho=10.7$ (d), corresponding to the motion on hyperboloids.
	The fixed points of the focus type are shown in blue, 
	while the saddle points and the saddle trajectories are plotted in red.
	}
    \label{fig:2}
\end{figure}

The system under consideration possesses the closed, pseudo-conservative trajectories only.
This is in sharp contrast to the other possible \cPT-symmetric configurations. For example, another way to add a non-Hermitian \cPT-symmetric part to the Hamiltonian is to introduce the pump-dissipation imbalance between the circular components, which corresponds to $H_1(\hat{\mathbf{s}})=-\gamma\hs{z}$. The resulting semiclassical dynamics of the dimer is characterized by disclosed blow-up trajectories and, therefore, by broken \cPT-symmetry \cite{barashenkov13}. While the dynamics is pseudo-conservative in our case, there is qualitative change in the topology of the manifold of trajectories with increasing dissipation imbalance $\gamma$. As shown in Fig.\ \ref{fig:2}, the trajectories reside on the spindle-shaped ellipsoids (prolate spheroids) for $\gamma<\ve$, and on the upper sheets of the two-sheet hyperboloids for $\gamma>\ve$. 

Another important feature is the change in the number of fixed points with increasing size of the ellipsoids or the hyperboloids. The size is controlled by the integral $\rho$. One can find the fixed points by setting the time derivatives to zero in Eqs.\ \eqref{XYZset}, that gives
\begin{subequations}\label{FixSet}
\begin{align}
    & y=0, \qquad z=\frac{\gamma\rho}{1-\gamma^{2}+x}, \label{FixSet-a} \\
    & (1-\gamma^{2}+x)^2x^2-\rho^{2}\left[(1+x)^{2}-\gamma^{2}\right]=0. \label{FixSet-b}
\end{align}
\end{subequations}

Standard analysis of the discriminant of Eq.\ \eqref{FixSet-b} reveals that this equation has two real roots for $\rho<\rho_c$, and four real roots for $\rho>\rho_c$, where the critical size $\rho_c$ is 
\begin{equation}\label{RhoC}
	\rho_c=[1+\gamma^{2/3}+\gamma^{4/3}]^{3/2}.
\end{equation}
This is exactly what happens in the case of weak dissipation imbalance ($\gamma<1$). There are two fixed points of the focus type for $0<\rho<\rho_c$, as shown in Fig.\ \ref{fig:2}(a). Two more fixed points appear due to the saddle-node bifurcation at $\rho=\rho_c$. The new focus and the saddle point initially appear at the same spin value, and they separate from each other with increasing $\rho$, see Fig.\ \ref{fig:2}(b). Similar bifurcation takes place in the hyperbolic case ($\gamma>1$). We note that $\rho$ can be negative in the hyperbolic case and one real root of Eq.\ \eqref{FixSet-b} is fictitious, so that there is only one focus below $\rho_c$ and three fixed points above it, see Fig.\ \ref{fig:2}(c,d). Exactly at $\gamma=1$, the spin trajectories resides on paraboloids, with one and three fixed points below and above $\rho_c=\sqrt{27}$, respectively. 

We note that there are two saddle trajectories related to the presence of a saddle point, see Fig.\ \ref{fig:2}(b,d). 
The time to go around the saddle trajectory is infinite. The periods of the spin orbits nearby are logarithmically large. 
Saddle trajectories correspond to degeneracy of two roots of the quartic polynomial in righthand-side of Eq.\ \eqref{DifEq4z}: $z_2=z_3=z_s$ with $z_1<z_s<z_4$.
In this case the modulus $m=1$, $\omega_1=z_s-z_1$, $\omega_4=z_4-z_s$, and $\cn({\omega}t,1)=1/\cosh({\omega}t)$. Eq.\ \eqref{Zsol} then gives two solutions that start from the saddle point $z_s$ at $t{\rightarrow}-\infty$, pass through either $z_1$ of $z_4$ at $t=0$, and return to the saddle point at $t{\rightarrow}+\infty$.

\emph{Discussion and conclusions.}
Generally speaking, non-resonantly excited polariton condensates are open systems obeying non-linear dissipative dynamics. Apart from typical wave phenomena, which sometimes interpreted using quantum language, their behavior is within the classical concepts. Truly quantum features, like entanglement, are not yet evidenced experimentally. To obtain the polariton dynamics the resembles that of the closed conservative systems, with possible quantum effects, it is necessary to detach the polariton condensates from the excitation spots of incoherent polariton reservoirs and decrease the shot noise produced by the harvest of polaritons by the condensate. 
The above implies working near the condensation threshold, and to avoid the instability related to passing above the threshold, it is preferential to have the distribution of dissipation rates in the system, so that one obtains a threshold range, where the external pump can be adjusted.

Distribution of dissipation rates depending on polarization of a trapped polariton condensate can be achieved by using weakly grated microcavities (Fig.\ \ref{fig:1}). We proposed to use special configuration, where the Hermitian Josephson splitting $\ve$ and the non-Hermitian splitting in dissipation rates $\gamma$ are introduced between different linear polarization pairs: horizontal-vertical and diagonal-antidiagonal, respectively. This way one can produce trapped polariton condensate subject to \cPT-symmetric Hamiltonian. 
Moreover, we have shown that, at least in semi-classical approximation, the \cPT-symmetry remains unbroken in the presence of polariton-polariton repulsive interaction. 

The dynamics of \cPT-symmetric polariton condensate is characterized by closed pseudo-conservative trajectories, which opens the perspective to use it as a coherent qubit. The analytical expressions for the trajectories are derived.
We have analyzed the bifurcations of the fixed points of this system and the crossover in the topology of the manifolds of trajectories.
At weak dissipation inbalance, $\gamma<\ve$, the spin of the condensate precesses on the spindle-shaped ellipsoids, which is topologically equivalent to the usual spin precession on a sphere. For the large dissipation inbalance, however, the precession takes place on hyperboloids and there is decrease in the number of fixed points by one. The dynamics of large spins is featured by the presence of saddle trajectories. In the vicinity of them, the motion is slow with large periods of precession. The rich spin dynamics of \cPT-symmetric trapped polariton condensates makes this platform promising for computational goals.

\emph{Acknowledgments.}
This work was supported in part by PAPIIT-UNAM Grant No.\ IN108524.

%

\end{document}